%% file: main.tex
\documentclass[10pt, conference, letterpaper]{IEEEtran}
\usepackage[yyyymmdd,hhmmss]{datetime}
\usepackage{multirow}

\usepackage{algpseudocode}
\usepackage{algorithm}
\usepackage{algorithmicx}
\usepackage{amsmath}
\usepackage{amssymb}
\usepackage{graphicx,caption}
\usepackage{subcaption}
\usepackage{color}
\usepackage{xspace}
\usepackage{float}
\usepackage{url}
\usepackage{lipsum}
\usepackage{verbatim}

\usepackage[printonlyused,withpage]{acronym}
\usepackage{xcolor}
\usepackage{amssymb}
\usepackage{graphicx}
\usepackage{bm}
\usepackage{cases}
\usepackage{cite}

\usepackage{fixme}
\fxsetup{
status=draft,
    layout=inline,
    theme=color,
    mode=multiuser
}

\setkeys{Gin}{draft=false}

\begin{document}

\title{A Study of Ultrawideband (UWB) Antenna Design for Cognitive Radio Applications}
\author{
Peshal B. Nayak, 
Sudhanshu Verma,
Preetam Kumar \\
Department of Electrical Engineering, \\
Indian Institute of Technology Patna 
}        
        

\maketitle
%

\input{secs/abstract}%
\input{secs/introduction}
\input{secs/cognitive_radio.tex}
\input{secs/uwbTxAndAntennaFeatures.tex}
\input{secs/antennaSystemsForCognitiveRadio.tex}
\input{secs/resultsAndDiscussion.tex}
\input{secs/conclusion.tex}

\bibliographystyle{unsrt}
\bibliography{ref_bib}
\end{document}

%% file: secs/abstract.tex
\begin{abstract}
Cognitive radio is rapidly shaping the future of wireless communications. Research on antenna design is very critical for the implementation of cognitive radio. A special antenna is required in cognitive radio for sensing and communicating. For the purpose of spectrum sensing, an Ultrawideband (UWB) antenna is being considered as a potential candidate by many experts. This paper provides a detailed discussion of the existing UWB spectrum sensing antenna designs for cognitive radio system. Simulation results for a promising cognitive radio antenna which provides a reconfigurable function in the range of 5-6 GHz have also been presented and shown to match closely with the measured results.\footnote{This document is an author's version of \cite{nayak2012ultrawideband}.}
\end{abstract}

%% file: secs/introduction.tex
\section{Introduction}
In recent times, there has been an explosive growth in wireless communications \cite{nayak2017multi, nayak2019ap, nayak2019modeling, nayak2019virtual, nayak2016performance, peshal2019modeling}. Also, it is expected that data traffic will double every year which will eventually result in the saturation of the dedicated spectrum. Currently, most spectrum bands have been allocated to licensed users. However, a lot of licensed bands such as those for TV broadcasting are underutilized resulting in spectrum wastage. As a result, the Federal Communications Commission (FCC) has been prompted to open licensed spectrum bands to unlicensed users through the use of Cognitive Radio (CR) technology. With the advent of 3G and 4G mobile communications, CR schemes have begun to receive a lot of attention \cite{tawk2011implementation}. 

Presently various research communities have different definitions of CR and its unique defining features. Some view it as primarily about dynamic spectrum sharing while some consider it as a device capable of cross-layer optimization. The possibilities for antenna to play an active role in system level performance are lost amid all these conceptions of CR. However, an antenna is the most important section of a CR system. Designing an antenna which carries out spectrum sensing as well as transmission is extremely difficult.

This paper provides a detailed discussion of the existing UWB sensing antennas in CR systems along with the simulation results for a recent promising antenna design. The paper is organized as follows. In section II, basic information on CR has been provided. We discuss UWB antenna features in Section III. Section IV provides details of the existing models of UWB antenna for CR systems. Simulation results for an Integrated UWB/ Reconfigurable slot antenna have been given in Section V. We give conclusion in the last section.

%% file: secs/cognitive_radio.tex
\section{Cognitive Radio}
According to the FCC, a cognitive radio is “a radio that can change its transmitter parameters based on the environment in which it operates”. Thus, for more efficient communication and spectrum use, a CR should be able to recognize spectrum availability and reconfigure itself accordingly \cite{force2002report}. Users who own the channel are termed as “primary” users (PUs) while the unlicensed users are called “secondary” users (SUs). These SUs need to continuously monitor the activities of the PUs to find unused frequency bands that can be utilized without any interference to the licensed services. This procedure is called as spectrum sensing and the unutilized bands are called as spectrum holes (SHs). 

Once a SH is found, the CR system should be able to adjust its parameters such as transmission power, carrier frequency, modulation strategy and transmission data rate, \cite{tawk2011comparison}, so that the unused frequency bands can be used by the SUs for transmission. These SUs can utilize the spectrum as long as the Quality of Service (QoS) is not compromised. However, channel conditions can change rapidly and so a SU has to continuously monitor all the licensed bands and keep looking for SHs. Spectrum sensing, therefore, plays a very crucial role in Cognitive Radio systems.

SDR which offers promise to increase spectrum usage efficiencies to users in a wide variety of applications is seen as an enabling technology for CR. Just like CR, SDR does not represent any specific concept. Wireless Innovation Forum, \cite{WirelessInnForum}, defines it as “a radio in which some or all the layer functions are software defined”.

\begin{figure}
    \centering
    \includegraphics[width=1\linewidth]{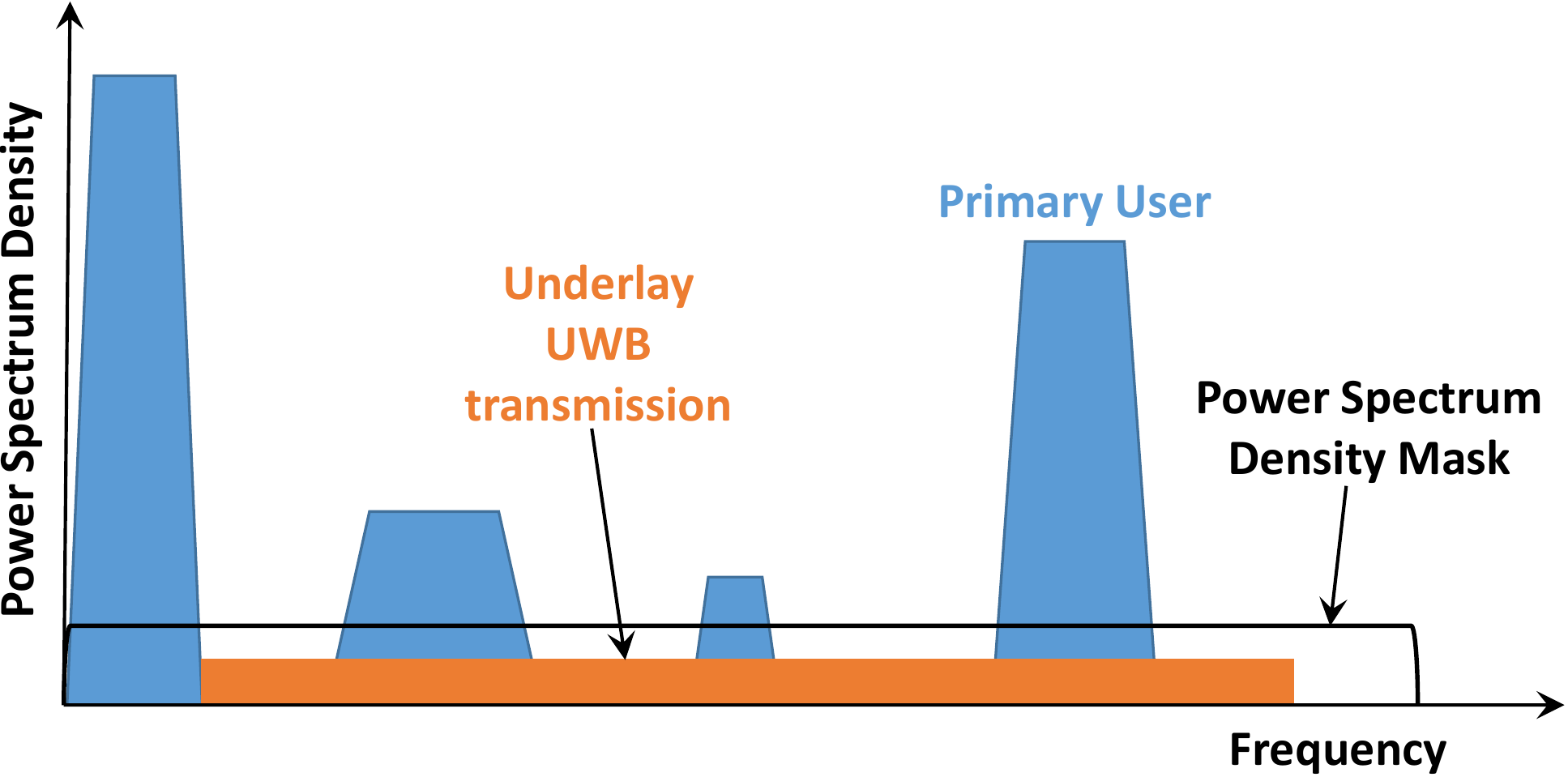}
    \caption{Underlay spectrum sharing approach}
    \label{fig:underlaySpectrum}
\end{figure}

\begin{figure}
    \centering
    \includegraphics[width=1\linewidth]{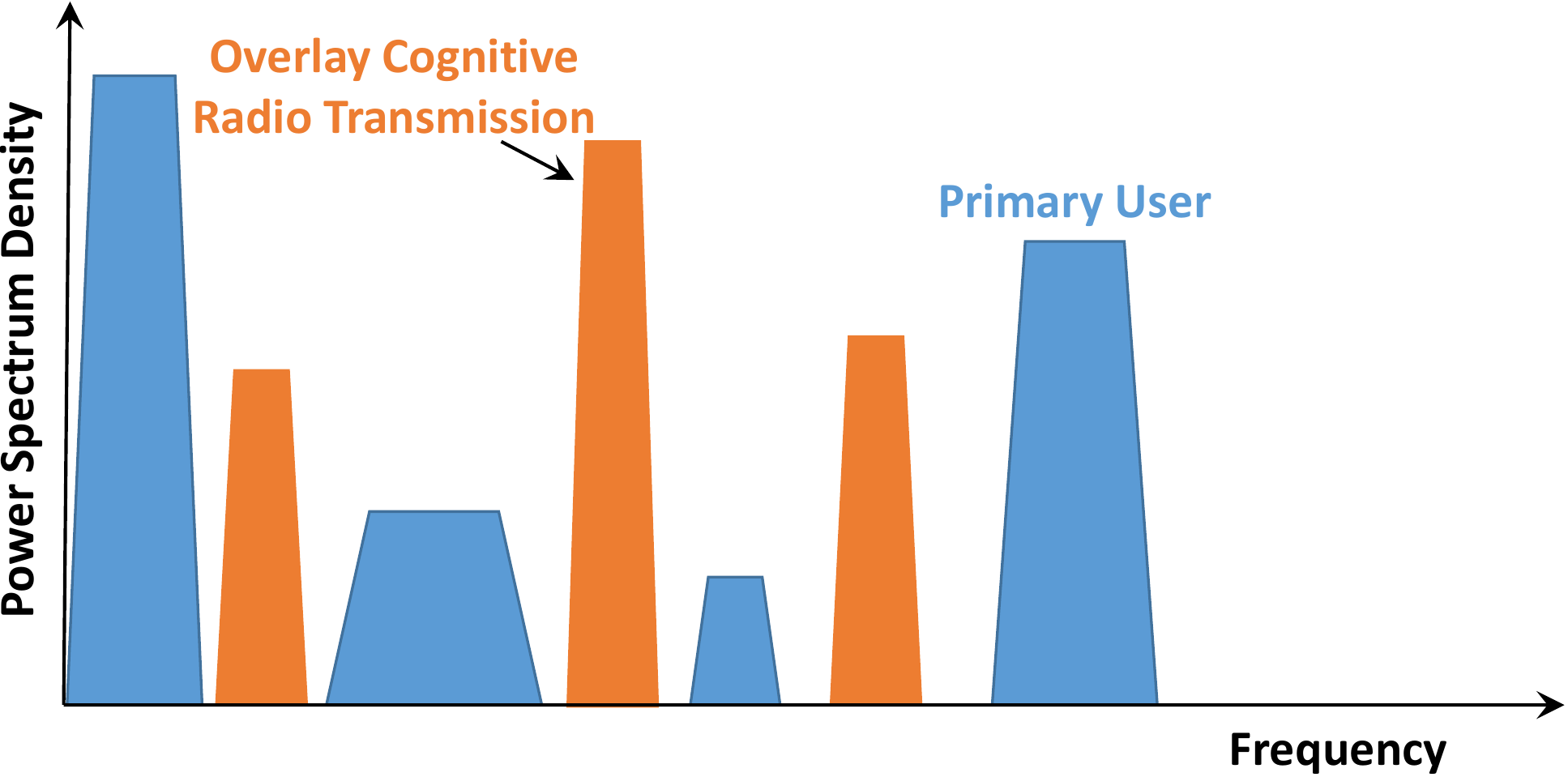}
    \caption{Overlay spectrum sharing approach}
    \label{fig:overlaySpectrum}
\end{figure}

As mentioned in \cite{tawk2011comparison}, many of the proposed dynamic spectrum sharing approaches (DSS) can be broadly classified into two categories viz. open sharing and hierarchical sharing. In open sharing, all users can simultaneously access the spectrum. There are, however, some limits on the transmit signal. In the hierarchical sharing model, licensed users are assumed to act sporadically in the frequency bands owned by them. As long as the QoS is not compromised, PUs may allow or lease their unutilized channels to the SUs. Three main paradigms have been considered in literature in this category. They are spectrum underlay, interweaving and overlay. In spectrum underlay, SUs may transmit simultaneously with PUs. But they should keep their transmission power below an interference margin or noise floor tolerated by the PUs. In spectrum interweave, SUs use appropriate mechanisms to locate SHs for their use. Thus, they should determine when and where they may transmit. On the other hand, cognitive users in spectrum overlay system already know the primary message. They use sophisticated implementation techniques to reduce the interference at the primary and secondary receivers. In short, they are allowed to utilize the SHs while avoiding or limiting collisions with primary transmission. The underlay and overlay approaches in the hierarchical model are illustrated in Fig.~\ref{fig:underlaySpectrum} and Fig.~\ref{fig:overlaySpectrum} respectively.

The development of spectrum sensing and spectrum sharing has facilitated the application of CR in many areas such as TV white spaces, cellular networks, military usage etc.  

%% file: secs/uwbTxAndAntennaFeatures.tex
\section{ULTRAWIDEBAND TRANSMISSION AND ANTENNA FEATURES}
In UWB transmission technique, pulses of very short duration of time are used across a very large frequency portion of the spectrum. These modulated high frequency pulses are of low power with duration of less than 1 nanosecond. As these waveforms are compressed in time, they have a very large bandwidth (greater than 500 MHz). UWB ranges from 3.1 GHz to 10.6 GHz. The role played by UWB antenna is that they are able to transmit these pulses accurately and efficiently. UWB transmissions are a part of the low background noise from the perspective of other communication systems. Therefore, it can be used without harmful interference to the primary communication systems and thus can be seen as an enabling technology for CR. It also has the capability of co-existing in the same temporal, spatial and spectral domains with other licensed/unlicensed radios. Thus, it is an underlay system and this is its most appealing property. Other useful features include multidimensional flexibility involving adaptable pulse shape, bandwidth data rate and transmit power. It also has low power consumption, resulting in low complexity transceivers which are necessary for portable devices. Low complexity also reduces the cost of the system. 

Following are the characteristic features that can be used to specify a UWB antenna \cite{adamiuk2012uwb}:
\begin{itemize}
    \item Beam width, side lobes.
    \item Polarization stability versus frequency.
    \item Constant gain versus frequency.
    \item High peak value of radiated impulse, high efficiency.
    \item Retain the signal bandwidth.
    \item Low dispersion of impulses in time domain.
    \item Low ringing in time domain.
    \item Constant group delay in frequency domain.
    \item Direction-independent impulse radiation, high fidelity.
    \item Stable phase centre over frequency.
\end{itemize}

Some redundancy is present in the above quantities, some of which result partly from each other. However, some of them are more convenient for time domain while others are more convenient for frequency domain.

%% file: secs/antennaSystemsForCognitiveRadio.tex
\section{ANTENNA SYSTEMS FOR COGNITIVE RADIO}

\begin{figure}
    \centering
    \includegraphics[width=1\linewidth]{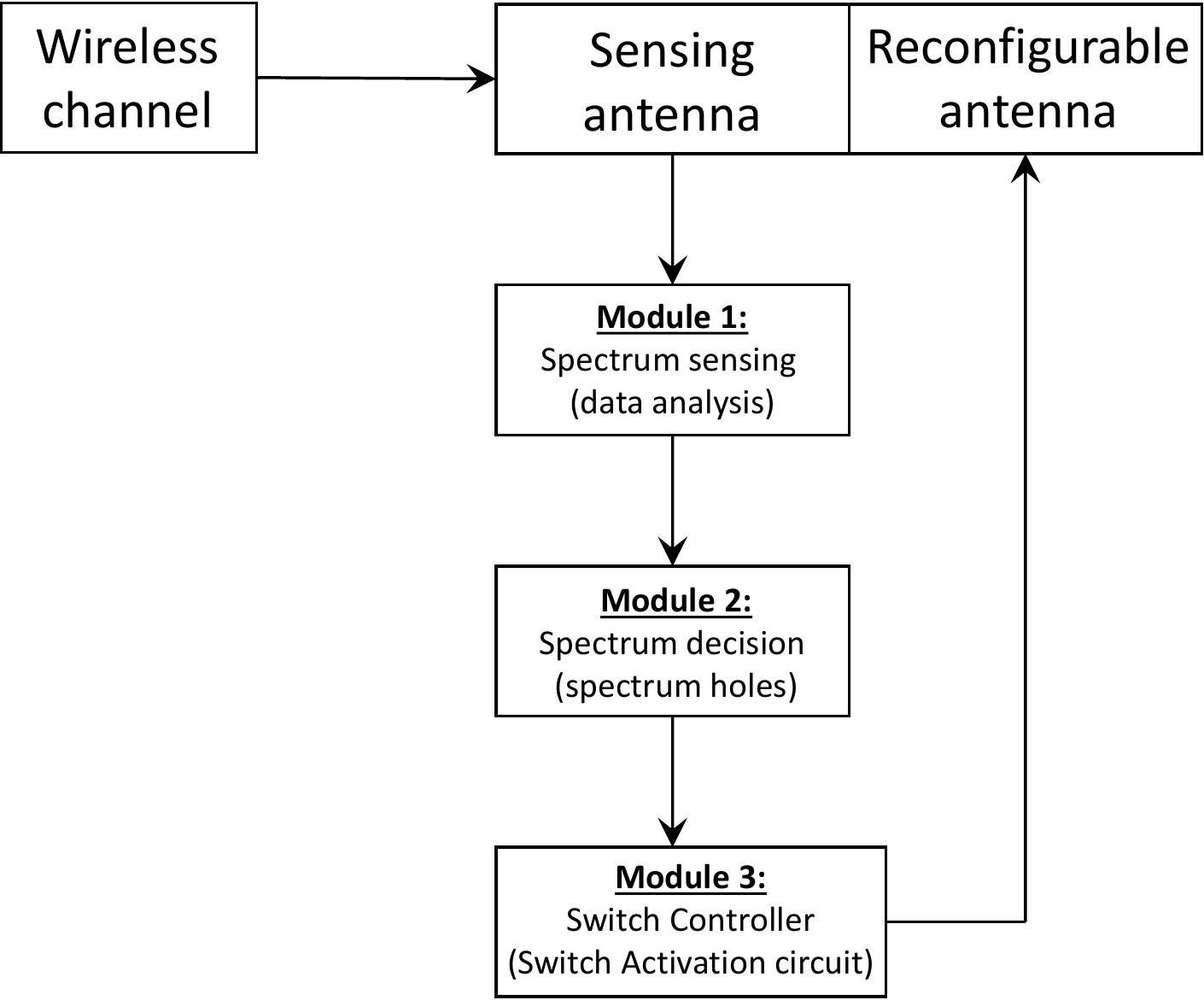}
    \caption{A generic cognitive radio workflow diagram}
    \label{fig:genericCogRadio}
\end{figure}

The basic RF structure of a CR system comprises of a “sensing antenna” and a “reconfigurable transmit receive antenna”. The task of the sensing antenna is to continuously monitor the wireless channel for unutilized frequency bands. The reconfigurable transmit/receive antenna then performs the required transmission in those channels \cite{tawk2009new}.  A generic cognitive radio workflow diagram is given in Fig.~\ref{fig:genericCogRadio}. Some experts have suggested that UWB antenna should be used for sensing operation. The sensing antenna communicates with the “spectrum sensing” module of the CR engine which continuously searches for unused frequency channels within the operating band of the sensing antenna. The information obtained by the “spectrum sensing” module is passed on to the “spectrum decision” module which determines the appropriate frequency band for communication. The switch controller then tunes the operating frequency of the reconfigurable antenna. 

The antennas used for CR networks are omnidirectional antennas with gains of 0dBi of higher \cite{narampanawe2011ultra}. They are primarily used for sensing and performing measurements. As a result, they are mounted outdoors for effective sensing. An example of a simple omni-directional antenna is the monopole antenna. Monopole antennas have the advantage of reduced physical dimensions but the drawback is that their bandwidths are relatively small \cite{hopperunderstanding}. If their bandwidth can be sufficiently improved, the monopole antenna can be a suitable design for CR applications.

The architecture for CR has not yet been standardized. Some reconfigurable antenna systems have been presented \cite{tawk2011comparison}. In antenna designs for interweave CR environment, the RF front end should be able to sense and search for SHs. The data acquired by sensing is then analyzed. The reconfigurable antenna is then tuned to transmit at specified frequencies. The sensing antenna structure in \cite{tawk2011comparison} covers the band from 3GHz to 11 GHz. In case of underlay CR environment, the capability to achieve UWB communication is desired due to restrictions on power level. To continuously transmit with low power in short distance communication, a UWB antenna is required. This antenna should have notches in its operating band to minimize interference between the primary and the secondary users.

As the demand for small sized high performance devices continuously increases, the space availability for mounting an antenna within a device becomes very limited. As a result, it has become necessary to miniaturize radiating element(s) of an antenna as well as integrate the reconfigurable narrow band and UWB antenna so that they share the same space \cite{kelly2010integrated, nayak2013multiband, nayak2014novel, nayak2013compact, endluri975low}. A technique based on integrating these two antennas into the same substrate has been proposed in \cite{ebrahimi2011integrated} and \cite{tawk2011implementation}. 

Recently various designs and architectures of a cognitive antenna have emerged. In most of the literature available on CR systems, a UWB antenna has been used as a sensing antenna. In \cite{tawk2011implementation}, a UWB egg shaped monopole antenna and five different narrow band patch radiators inside a circular section have been printed on the same substrate. By physically rotating the circular part via a stepper motor, the operating frequency of this antenna can be adjusted. At each rotation step, a frequency band is obtained by feeding an individual patch radiator. A stepper motor requires more space, thereby increasing the complexity and the cost of the antenna.

In \cite{ebrahimi2011integrated}, as a narrow band antenna, a planar inverted F-resonator is printed on the reverse side of a coplanar waveguide (CPW)-fed UWB monopole antenna. It utilizes the radiator of the UWB as its own ground plane. A matching circuit has been used to tune this antenna for three different regions centered at 4, 8 and 10 GHz which has increased the antenna complexity and size.

In another technique, sensing and communicating antennas are realized by switching between a narrow band antenna and a UWB resonator \cite{zamudio2011integrated, hamid2011vivaldi, kelly2009reconfigurable}. In this case, a single terminal feeds the antenna structure. This method is achieved by two ways:
\begin{itemize}
    \item Incorporating a band pass structure inside a UWB antenna \cite{zamudio2011integrated, hamid2011vivaldi}.
    \item Changing the structure of the antenna radiator or the ground plane via switches \cite{kelly2009reconfigurable}.
\end{itemize}

A reconfigurable band pass filter is integrated with a UWB antenna in \cite{zamudio2011integrated}. The reconfigurability is based on incorporating nine switches within the defected micro-strip structure (DMS) band pass filter.

In \cite{erfani2012design}, an integrated elliptical monopole antenna with reconfigurable slot radiator on the same substrate has been introduced. The antenna configuration is shown in Fig. 4. The isolation between the narrowband and the UWB has been reduced to better than -16dB by folding the slot resonator current distribution using a balanced stub inside the slot. A small port-small size antenna for cognitive radio applications is presented in \cite{al2011uwb}. The antenna is based on UWB design and has a reconfigurable band pass filter integrated in its feed line. Electronic switches are incorporated on the filter to activate/ deactivate it and to control its band pass frequency.    

%% file: secs/resultsAndDiscussion.tex
\section{RESULTS AND DISCUSSION}
\begin{figure}
    \centering
    \includegraphics[width=1\linewidth]{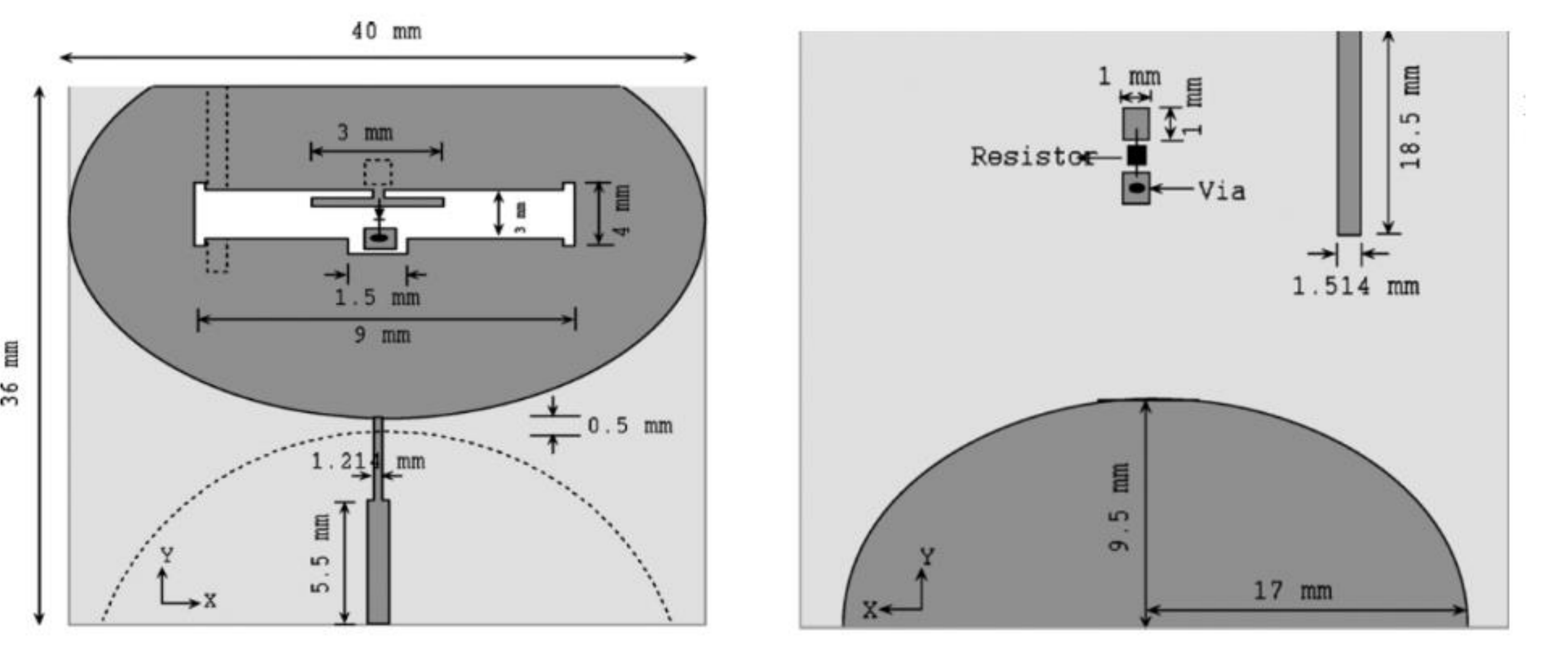}
    \caption{Configuration of the UWB/reconfigurable narrowband antenna (a)Top view (b)Bottom view \cite{erfani2012design}}
    \label{fig:AntenaViews}
\end{figure}

\begin{figure}
    \centering
    \includegraphics[width=1\linewidth]{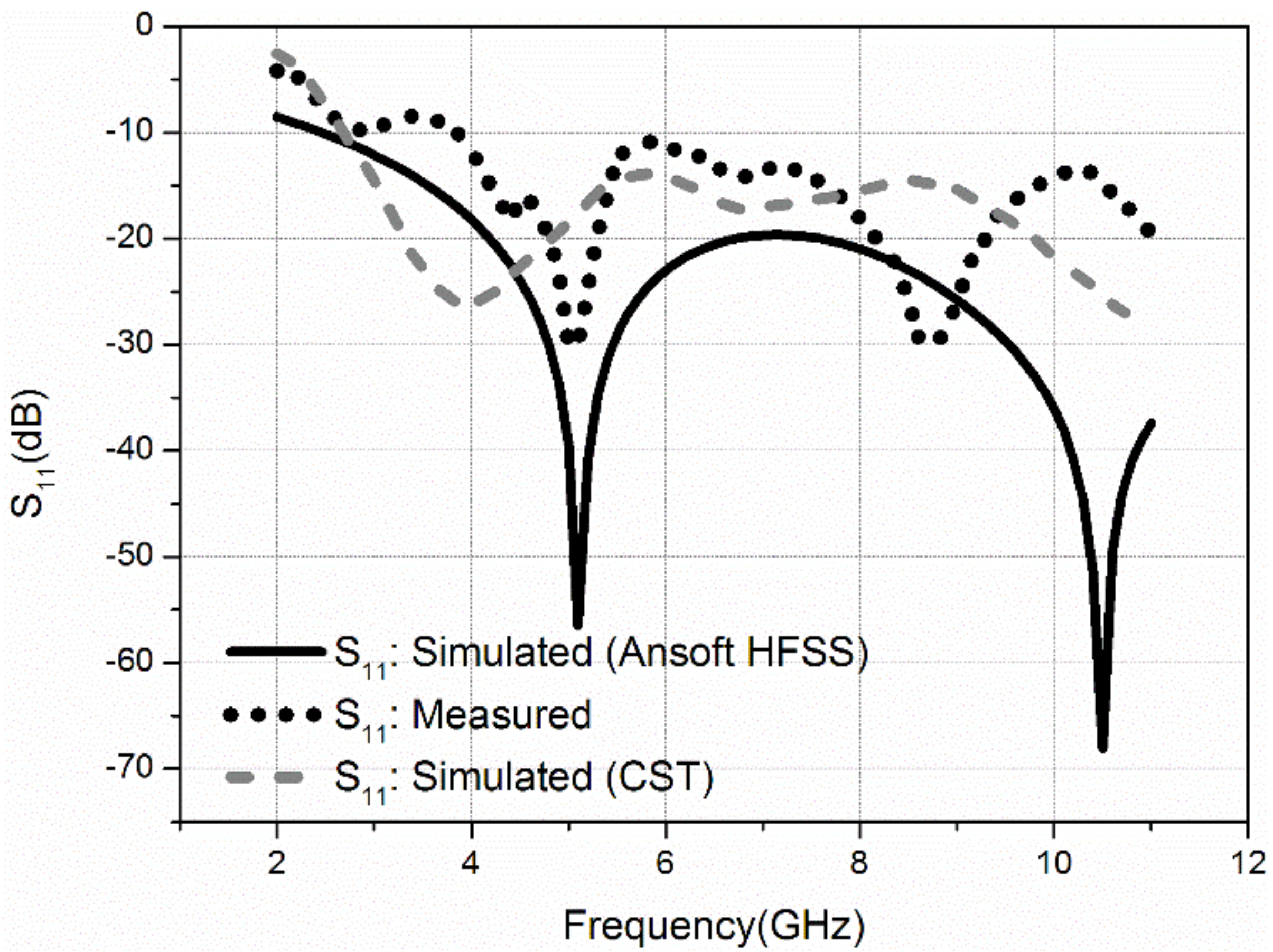}
    \caption{Measured and simulated reflection coefficient of sensing antenna}
    \label{fig:Results}
\end{figure}

The antenna configuration proposed in \cite{erfani2012design} has been simulated to validate its sensing operation for cognitive radio. Figure~\ref{fig:AntenaViews} shows the antenna configuration and the dimensions as discussed in \cite{erfani2012design} while the simulation results have been shown in Fig.~\ref{fig:Results}. An elliptical disc fed by a microstrip line has been printed on a 40x36 $mm^{2}$ RO4350B substrate ($\epsilon_{r}$ = 3.48, tang $\gamma$ = 0.0037) with thickness of 0.662 mm. The major and minor radii of the partial ellipse etched on the bottom layer are $R_{x}$ = 17 mm and $R_{y}$ = 9.5 mm. This has been used as a ground plane. A step-fed matching technique is used in the feeding line to control the input impedance across the desired band. A symmetric stub is used inside the slot to reduce the effective length of the resonant slot by folding the slot current distribution. The varactor diode placed across the slot is used to reconfigure the operating frequency of the narrowband antenna. When biased (5 $V_{dc}$), the varactor diode resonates at 6 GHz. This resonant frequency is decreased by reducing the bias voltage. The varactor diode is accommodated inside the slot by creating an isolated pad. 

This antenna configuration has been simulated with Ansoft High Frequency Structure Simulation (HFSS) software, version 13. In \cite{erfani2012design}, Computer Simulation Technology (CST) software has been used for the simulation of the proposed antenna. Those results have also been shown in the Fig. 5. It can be clearly seen that our simulation results show a better match with the measured values of S11 as compared to those obtained by CST simulation. Figure 5 also shows that the operating band ranges from 2.5 to 11 GHz which covers the entire UWB band (3.1 to 10.6 GHz). Other results stated in \cite{erfani2012design} regarding the performance of the antenna design have also been verified. 

%% file: secs/conclusion.tex
\section{Conclusion}
In this paper, the application of UWB antenna as a sensing antenna has been discussed. A UWB antenna integrated with a reconfigurable antenna can be used for efficient sensing and subsequent use of unutilized frequency bands in Cognitive Radio systems. Integrating techniques to reduce the size of the antenna have also been discussed and highlighted. Important features of various CR antenna systems which use UWB antenna for sensing have been reported. Results for a recently proposed UWB antenna for sensing application in CR have been verified through simulation.  